%% file: fairnessAwareRankingInSearchAndRecommendationSystems.tex
\documentclass[sigconf]{acmart}
\renewcommand\footnotetextcopyrightpermission[1]{} 

\usepackage{booktabs}

\usepackage{url,bm}
\usepackage{latexsym}

\usepackage{algorithm}
\usepackage[noend]{algorithmic}
\usepackage{multirow}
\usepackage{xcolor}

\usepackage{etoolbox}
\newtoggle{conf}
\toggletrue{conf}

\newenvironment{compact_enum}{
\begin{itemize}
  \setlength{\itemsep}{0pt}
  \setlength{\parskip}{1pt}
  \setlength{\parsep}{0pt}
  \setlength{\itemindent}{-5pt}
}{\end{itemize}}

\newenvironment{compact_numbered_enum}{
\begin{enumerate}
  \setlength{\itemsep}{0pt}
  \setlength{\parskip}{0pt}
  \setlength{\parsep}{0pt}
  \setlength{\itemindent}{-5pt}
}{\end{enumerate}}

\newcommand{\cut}[1]{}

\newcommand{\taur}{{\tau_r}}
\newcommand{\taurk}{{\tau_{r}^{k}}}
\newcommand{\tauri}{{\tau_{r}^{i}}}

\newcommand{\pidealtildep}{{p_{q, r, a_i}}}
\newcommand{\porigtildep}{{p_{\taur, r, a_i}}}
\newcommand{\porigktildep}{{p_{\taurk, r, a_i}}}

\usepackage{mathtools}
\DeclarePairedDelimiter\ceil{\lceil}{\rceil}
\DeclarePairedDelimiter\floor{\lfloor}{\rfloor}

\copyrightyear{2019}
\acmYear{2019} 
\setcopyright{acmcopyright}
\acmConference{KDD 2019}{}{Aug 4--8, 2019, Anchorage, AK.}
\acmPrice{15.00}
\acmDOI{http://dx.doi.org/xx.yyyy/zzzzzzz.zzzzzzz}
\acmISBN{ISBN 123-4-5678-9012-3/34/56}

\begin{document}

\title{Fairness-Aware Ranking in Search \& Recommendation Systems\\ with Application to LinkedIn Talent Search}

\author{Sahin Cem Geyik, Stuart Ambler, Krishnaram Kenthapadi}
\affiliation{
  \institution{LinkedIn Corporation, USA}
}
\renewcommand{\shortauthors}{S. Geyik et al.}

\begin{abstract}
\input{abstract}
\end{abstract}


\maketitle

\input{intro}
\input{evaluatingBias}

\input{algorithms}
\input{results}
\input{relatedwork}

\input{conclusion}

\begin{acks}
\input{acknowledgments}
\end{acks}

{\scriptsize
\bibliographystyle{abbrv} 
\bibliography{paper}
}

\clearpage

\appendix
\input{appendix}

\end{document}

%% file: abstract.tex
We present a framework for quantifying and mitigating algorithmic bias in mechanisms designed for ranking individuals, typically used as part of web-scale search and recommendation systems. We first propose complementary measures to quantify bias with respect to protected attributes such as gender and age. We then present algorithms for computing fairness-aware re-ranking of results. For a given search or recommendation task, our algorithms seek to achieve a desired distribution of top ranked results with respect to one or more protected attributes. We show that such a framework can be tailored to achieve fairness criteria such as \emph{equality of opportunity} and \emph{demographic parity} depending on the choice of the desired distribution. We evaluate the proposed algorithms via extensive simulations over different parameter choices, and study the effect of fairness-aware ranking on both bias and utility measures. We finally present the online A/B testing results from applying our framework towards representative ranking in LinkedIn Talent Search, and discuss the lessons learned in practice. Our approach resulted in tremendous improvement in the fairness metrics (nearly three fold increase in the number of search queries with representative results) without affecting the business metrics, which paved the way for deployment to 100\% of \emph{LinkedIn Recruiter} users worldwide. Ours is the first large-scale deployed framework for ensuring fairness in the hiring domain, with the potential positive impact for more than 630M LinkedIn members.\let\thefootnote\relax\footnote{\\ ~ \\ ~  \large \textbf{This paper has been accepted for publication at ACM KDD 2019.}}

%% file: intro.tex
\section{Introduction}\label{sec:intro}
Ranking algorithms form the core of search and recommendation systems for several applications such as hiring, lending, and college admissions. Recent studies show that ranked lists produced by a biased machine learning model can result in systematic discrimination and reduced visibility for an already disadvantaged group~\cite{dwork_2012, hajian_2016_tutorial, pedreschi_2009}
(e.g., disproportionate association of higher risk scores of recidivism with minorities~\cite{angwin_2016}, over/under-representation and racial/gender stereotypes in image search results~\cite{kay_2015}, and incorporation of gender and other biases as part of algorithmic tools~\cite{bolukbasi_2016, caliskan_2017}). One possible reason is that machine learned prediction models trained on datasets that exhibit existing societal biases end up learning them and can reinforce such bias in their results, potentially even amplifying the effect. 

In this paper, we present a framework for quantifying and mitigating algorithmic bias in systems designed for ranking individuals. Given fairness requirements expressed in terms of a desired distribution over protected attribute(s) (e.g., gender, age, or their combination), we propose algorithms for re-ranking candidates scored/returned by a machine learned model to satisfy the fairness constraints. Our key contributions include:
\begin{compact_enum}
\item Proposal of fairness-aware ranking algorithms towards mitigating algorithmic bias. Our methodology can be used to achieve fairness criteria such as \emph{equality of opportunity}~\cite{hardt_2016} and \emph{demographic parity}~\cite{dwork_2012} depending on the choice of the desired distribution over protected attribute(s).
\item Proposal of complementary measures for quantifying the fairness of the ranked candidate lists.
\item Extensive evaluation of the proposed algorithms via simulations over a wide range of ranking scenarios and attributes with different cardinalities (possible number of values).
\item Online A/B test results of applying our framework for achieving representative ranking in LinkedIn Talent Search, and the lessons learned in practice. Our approach resulted in tremendous improvement in the fairness metrics (nearly three fold increase in the number of search queries with representative results) without statistically significant change in the business metrics, which paved the way for deployment to 100\% of \emph{LinkedIn Recruiter} users worldwide.
\end{compact_enum}

The rest of the paper is organized as follows. We first provide measures for evaluating bias and fairness in ranked lists in \S\ref{sec:evaluatingBias}. Next, we present fairness-aware re-ranking algorithms in \S\ref{sec:algorithms}, followed by their extensive evaluation and results from deployment in LinkedIn Talent Search in \S\ref{sec:results}. We discuss related work as well as a comparison of our approach to previous work on fairness-aware ranking in \S\ref{sec:background}. We conclude the paper and present future work in \S\ref{sec:conclusion}.

%% file: evaluatingBias.tex
\section{Measuring Algorithmic Bias}\label{sec:evaluatingBias}
We first discuss the intuition underlying our bias measurement approach, and then present measures for quantifying bias in ranking that are complementary to each other.

\subsection{Intuition Underlying Bias Quantification} \label{subsec:intuition}
Our bias measurement and mitigation approach assume that in the ideal setting, the set of top ranked results for a search or recommendation task should follow a desired distribution over a protected attribute such as gender or age. This desired distribution can be computed in many ways including, but not limited to, adhering to 
the corresponding distribution over a baseline population,
a legal mandate, or a voluntary commitment (e.g.,~\cite{ec2018, useeoc2017, verge_2010}). Note that our framework allows fairness-aware re-ranking over multiple attributes by considering the cross-product of possible values, e.g., adhering to a desired distribution over all possible (gender, age group) pairs. 
As we discuss in \S\ref{sec:mappingtofairnessnotions}, we can achieve fairness criteria such as \emph{equal opportunity}~\cite{hardt_2016} and \emph{demographic parity}~\cite{dwork_2012} depending on the choice of the desired distribution.

\subsection{Measures for Bias Evaluation} \label{subsec:metrics}
We next describe measures for evaluating bias in recommendation and search systems. We use the notations listed in Table~\ref{tab:notations}.

\begin{table}[!ht]
\caption{Key Notations}
\vspace{-10pt}
\small
\begin{tabular}{c|p{5.5cm}}
\hline\hline
\textbf{Notation} & \textbf{Represents} \\ \hline\hline
$r$ & A search request or a recommendation task \\ \hline
$A = \{ a_1, \ldots, a_l \}$ & Set of disjoint protected attribute values (each  candidate has exactly one value in $A$); Note that we denote the attribute value for candidate $x$ as $A(x)$, by abuse of notation. \\ \hline
$\taur$ &  Ranked list of candidates for $r$; $\taur[j]$ denotes $j^{th}$ candidate; $\taurk$ denotes the first $k$ candidates in $\taur$\\ \hline
$\pidealtildep$ & Desired proportion of candidates with attribute value $a_i$ that should be in the ranked list \\ \hline
$\porigtildep$ & Proportion of candidates in $\taur$ with value $a_i$, i.e., $\frac{\{ x \in \taur | A(x) = a_i \}}{|\taur|}$ \\ \hline
\hline
\end{tabular}
\label{tab:notations}
\end{table}

\subsubsection{Measure based on Top-k Results} \label{subsubsec:skewatk}
Our first measure computes the extent to which the set of top $k$ ranked results for a search or recommendation task differ over an attribute value with respect to the desired proportion of that attribute value.
\begin{definition}
\label{def:skew_at_k}
Given a ranked list $\taur$ of candidates for a search request $r$, the \emph{skew} of $\taur$ for an attribute value $a_i$ is:
\small
\begin{equation}
\label{eq:skew_at_k}
Skew_{a_i}@k(\taur) = \log_e\bigg(\frac{\porigktildep}{\pidealtildep}\bigg) ~ .
\end{equation}
\normalsize
\end{definition}
In other words, $Skew_{a_i}@k$ is the (logarithmic) ratio of the proportion of candidates having the attribute value $a_i$ among the top $k$ ranked results to the corresponding desired proportion for $a_i$. A negative $Skew_{a_i}@k$ corresponds to a lesser than desired representation of candidates with value $a_i$ in the top $k$ results, while a positive $Skew_{a_i}@k$ corresponds to favoring such candidates. We utilize the $\log$ to make the skew values symmetric around origin with respect to ratios for and against a specific attribute value $a_i$. For example, the ratio of the proportions being $2$ or $\frac{1}{2}$ corresponds to the same skew value in magnitude, but with opposite signs. Note that the calculation might need some adjustment to prevent a case of divide-by-zero or $\log(0)$.

Consider the gender attribute (with values \{$a_1 = $ male, $a_2 =$ female\}) as an example. Suppose that, for a given search task, the desired proportions are obtained based on the set of qualified candidates which consists of $32K$ males and $48K$ females ($80K$ total, hence desired ratios are $p_{q, r, \textrm{male}}=0.4$ and $p_{q, r, \textrm{female}}=0.6$). If the set of top 100 ranked results for this task consists of 20 males and 80 females, then, $Skew_{male}@100 = \log_e(\frac{20}{100} / \frac{32K}{80K}) = \log_e(0.5) \approx -0.3$.

$Skew_{a_i}@k$ measure is intuitive to explain and easy to interpret. In the above example, we can infer that males are represented 50\% less than the desired representation. However, $Skew_{a_i}@k$ has the following disadvantages. (1)  It is defined for a single attribute value, and hence we may need to compute the skew value for all possible values of the protected attribute. (2) It depends on $k$ and has to be computed for different $k$ values to fully understand the extent of the bias. While certain choices of $k$ may be suitable based on the application (e.g., $k = 25$ may be meaningful to measure skew in the first page of results for a search engine that displays 25 results in each page), a measure that takes into account all candidates in a ranked list may be desirable to provide a more holistic view of fairness.

To deal with the first problem above, we introduce two more measures which give a combined view of $Skew@k$ measure:
\begin{itemize}
\item \emph{MinSkew@k}: For a search request $r$, \emph{MinSkew@k} provides the minimum skew among all attribute values,
\small
\begin{equation}
\label{eq:minskew_at_k}
MinSkew@k(\taur) = min_{a_i \in A} Skew_{a_i}@k(\taur) ~ .
\end{equation}
\normalsize
\item \emph{MaxSkew@k}: For a search request $r$, \emph{MaxSkew@k} provides the maximum skew among all attribute values,
\small
\begin{equation}
\label{eq:maxskew_at_k}
MaxSkew@k(\taur) = max_{a_i \in A} Skew_{a_i}@k(\taur) ~ .
\end{equation}
\normalsize
\end{itemize}
MinSkew and MaxSkew have the following interpretation. MinSkew signifies the \emph{worst disadvantage in representation} given to candidates with a specific attribute value while MaxSkew signifies the \emph{largest unfair advantage} provided to candidates with an attribute value. Since both $\sum \porigktildep = 1$ and $\sum \pidealtildep = 1$, it follows that for any ranked list, and for any $k$, $MinSkew@k \leq 0$ and $MaxSkew@k \geq 0$.

Next, we present a ranking measure that addresses the second problem with skew measure as presented above.

\subsubsection{Ranking Measure}\label{subsubsec:ndkl}
Several measures for evaluating the fairness of a ranked list have been explored in the information retrieval literature \cite{yang_2017}. In this paper, we adopt a  ranking bias measure based on Kullback-Leibler (KL) divergence~\cite{kullback_1951}. Let $D_{\tauri}$ and $D_{r}$ denote the discrete distribution assigning to each attribute value in $A$, the proportion of candidates having that value, over the top $i$ candidates in the given ranked list $\taur$ and over the desired distribution respectively. Given these two distributions, we compute the KL-divergence and then obtain a normalized discounted cumulative variant, similar to~\cite{yang_2017}. This measure is non-negative, with a larger value denoting greater divergence between the two distributions. It equals 0 in the ideal case of the two distributions being identical for each position $i$.
\begin{definition}
\label{def:ndkl}
Given a ranked list $\taur$ of candidates for a search request $r$, the \emph{normalized discounted cumulative KL-divergence} (NDKL) of $\taur$ is:
\small
\begin{equation}
\label{eq:ndkl}
NDKL(\taur) = \frac{1}{Z} ~ \sum_{i=1}^{|\taur|} \frac{1}{\log_2(i + 1)} d_{KL}(D_{\tauri} || D_{r})~,
\end{equation}
\normalsize
where, $d_{KL}(D_1 || D_2) = \sum_j D_1(j) \log_e \frac{D_1(j)}{D_2(j)}$ is the KL-divergence of distribution $D_1$ with respect to distribution $D_2$ and \smaller{$Z = \sum_{i=1}^{|\taur|} \frac{1}{\log_2(i + 1)}$}.
\end{definition}
Note that $d_{KL}(D_{\tauri} || D_{r})$ corresponds to a weighted average of $Skew@i$ over all attribute values.
While having the benefit of providing a single measure of bias over all attribute values and a holistic view over the whole ranked list, the NDKL measure has the following disadvantages. (1) It cannot differentiate between bias of equal extent, but in opposite directions. For example, given an equal desired proportion of males and females (i.e., $p_{q, r, \textrm{male}}=p_{q, r, \textrm{female}}=0.5$), NDKL would be the same irrespective of whether males or females are being under-represented in the top ranked results by the same extent. Thus, the measure does not convey which attribute value is being unfairly treated (Skew measure is more suitable for this). (2) It is not as easy to interpret as the skew measure.

%% file: algorithms.tex
\section{\large{Fairness-aware Ranking Algorithms}}\label{sec:algorithms}
We next present a discussion of the desired properties when designing fair ranking algorithms, followed by a description of our proposed algorithms.

\subsection{Discussion of Desired Properties}
As presented in \S\ref{sec:evaluatingBias}, we assume that for each attribute value $a_i$, it is desirable for a fair ranking algorithm to include candidates possessing $a_i$ with a proportion as close as possible to $p_{q,r,a_i}$ (for brevity, we also use the term $p_{a_i}$ to mean the desired proportion of candidates possessing attribute value $a_i$). While one can argue that for a representation proportion of $\porigtildep > p_{q,r,a_i}$, we are still ``fair'' to $a_i$, a model that achieves such a recommendation proportion could cause unfairness to some other $a_j \neq a_i$, since $\sum_{a \in A} p_{q,r, a} = \sum_{a \in A} p_{\taur, r, a} = 1$. This is the case because the attribute values are disjoint, i.e., each candidate possesses exactly one value of a given attribute.

Furthermore, it is desirable for the representation criteria to be satisfied over top-k results for all $1 \le k \le |\taur|$, since presenting a candidate earlier vs. later in the ranking could have a significant effect on the response of the user~\cite{joachims_2005}.
Thus, we would like the ranked list of candidates to satisfy the following desirable properties:
\small
\begin{equation}
\label{eq:ranking_objective_1}
\forall k \leq |\taur| ~ \& ~ \forall a_i \in A, ~ count_k(a_i) \leq \ceil{p_{a_i} \cdot k} ~, ~ and,
\end{equation}
\begin{equation}
\label{eq:ranking_objective_2}
\forall k \leq |\taur| ~ \& ~ \forall a_i \in A, ~ count_k(a_i) \geq \floor{p_{a_i} \cdot k} ~,
\end{equation}
\normalsize
where $count_k(a_i)$ denotes the number of candidates with attribute value $a_i$ among the top k results.  Among the two conditions above, Eq.~\ref{eq:ranking_objective_2} is more important for fairness purposes, since it guarantees a minimum representation for an attribute value (Eq.~\ref{eq:ranking_objective_1} helps to ensure that disproportionate advantage is not given to any specific attribute value, since this could cause disadvantage to other attribute values). We next define a notion of \emph{(in)feasibility} for a ranking algorithm in terms of fairness.
\begin{definition}
\label{def:feasibility}
A ranking algorithm is infeasible if:
\small
\begin{equation}
\label{eq:feasibility}
\exists ~ r ~~s.t.~~ \exists ~ k \leq |\taur| ~\& ~ a_i \in A, count_k(a_i) < \floor{p_{a_i} \cdot k} ~.
\end{equation}
\normalsize
\end{definition}
{\noindent This} means that there is at least one search request $r$, such that the generated ranking list $\taur$ breaks the condition $count_k(a_i) \geq \floor{p_{a_i} \cdot k}$ for at least one $k$. We define the following measures to quantify the extent of infeasibility.
\begin{itemize}
\item \emph{InfeasibleIndex:} is defined as the number of indices $k \leq |\taur|$ for which (\ref{eq:ranking_objective_2}) is violated.
\small
\begin{equation}
\label{eq:infeasible_index}
\textrm{InfeasibleIndex}_{\taur} = \sum_{k \leq |\taur|} 1{(\exists a_i \in A, ~s.t. ~ count_k(a_i) < \floor{p_{a_i} \cdot k})}.
\end{equation}
\normalsize
While this value depends on the size of the ranked list $\taur$, it can be normalized if needed.
\item \emph{InfeasibleCount:} is defined as the number of (attribute value $a_i$, index $k$) pairs for which (\ref{eq:ranking_objective_2}) is violated.
\small
\begin{equation}
\label{eq:infeasible_count}
\textrm{InfeasibleCount}_{\taur} = \sum_{k \leq |\taur|} \sum_{a_i \in A} 1{(count_k(a_i) < \floor{p_{a_i} \cdot k})} ~.
\end{equation}
\normalsize
While this value depends on the size of the ranked list $\taur$, as well as the number of possible attribute values ($|A|$), it can again be normalized.
\end{itemize}
Next, we present our proposed set of algorithms for obtaining fair re-ranked lists. Note that the proposed algorithms assume that there are enough candidates for each attribute value, which may not always be the case in search and recommendation systems.
However, it would be easy to modify all the proposed algorithms to have a fallback mechanism to choose another candidate from the next-best attribute value (for fairness purposes). Finally, to avoid repetition, we have listed the combined set of inputs and outputs for all the algorithms in Table~\ref{table:inputoutputs}.

\begin{table}
\centering
\caption{\small Collective Inputs and Outputs of Algorithms~\ref{alg:detgreedy} through \ref{alg:detconstsort}}
\vspace{-10pt}
\small
\begin{tabular}{c|l} \hline \hline
\multirow{9}{*}{\textbf{Inputs}} 		& $\bm{a}$: Possible attribute values indexed as $a_i$, with each \\
													& attribute value having $n$ candidates with scores $s_{a_i,\cdot}$.\\
													& Candidate list for each attribute value is assumed to be\\
													& ordered by decreasing scores, i.e., for $j \geq 0$, $a_{i,j}$ refers to \\
													& $j^{th}$ element of attribute value $a_{i}$, with score $s_{a_i,j}$.\\
													& $\forall k, l:  0 \leq k \leq l \iff s_{a_i,k} \ge s_{a_i,l}$\\ \cline{2-2}
													& $\bm{p}$: A categorical distribution where $p_{a_i}$ indicates the \\
													& desired proportion of candidates with attribute value $a_i$\\
													\cline{2-2}
													& \textbf{$k_{max}$}: Number of desired results\\
													\hline
\textbf{Output} 								& An ordered list of attribute value ids and scores  \\ \hline\hline
\end{tabular}
\label{table:inputoutputs}
\end{table}

\subsection{Ranking Algorithms}

\subsubsection{Baseline algorithm with no mitigation (Vanilla)}
Our baseline ranking approach orders candidates in the descending order of score assigned by the ML model.

We next present four deterministic algorithms towards the goal of satisfying the conditions given in Eq.~\ref{eq:ranking_objective_1} and Eq.~\ref{eq:ranking_objective_2}.

\subsubsection{Score maximizing greedy mitigation algorithm (DetGreedy)}
Deterministic Greedy (DetGreedy) algorithm (Alg.~\ref{alg:detgreedy}) works as follows: If there are any attribute values for which the minimum representation requirement (Eq.~\ref{eq:ranking_objective_2}) is about to be violated, choose the one with the highest next score among them. Otherwise, choose the attribute value with the highest next score among those that have not yet met their maximum requirements (Eq.~\ref{eq:ranking_objective_1}).  At the end of each iteration in Alg.~\ref{alg:detgreedy}, $counts[a_i]$ maintains the number of candidates with attribute value $a_i$ included in the top $k$ results ($count_k(a_i)$) and $s_{a_i, counts[a_i]}$ denotes the score of the next best candidate (not yet shown) from attribute value $a_i$.

\begin{algorithm}[t]
\caption{\small{Score Maximizing Greedy Mitigation Algorithm (DetGreedy)}}
\label{alg:detgreedy}
\small
\begin{algorithmic}[1]

\STATE {\bf foreach} $a_i ~ \in ~ a$, $counts[a_i] := 0$
\STATE rankedAttList $:= [ ]$; ~ ~ rankedScoreList $:= [ ]$

\FOR {$k \in \{1, \ldots, k_{max} \}$}
	\STATE belowMin $:= \{ a_i: counts[a_i] < \floor{k \cdot p_{a_i}} \}$
	\STATE belowMax $:= \{ a_i: counts[a_i] \geq \floor{k \cdot p_{a_i}} \textrm{ and } counts[a_i] < \ceil{k \cdot p_{a_i}} \}$
	\IF {belowMin $\neq$ $\emptyset$}
		\STATE nextAtt $:=$ argmax$_{a_i \in \textrm{belowMin}} ~ s_{a_i, counts[a_i]}$
	\ELSE
		\STATE nextAtt $:=$ argmax$_{a_i \in \textrm{belowMax}} ~ s_{a_i, counts[a_i]}$
	\ENDIF
	\STATE rankedAttList[k] $:=$ nextAtt
	\STATE rankedScoreList[k] $:= s_{nextAtt, ~ counts[nextAtt]}$
	\STATE counts[nextAtt]$++$
\ENDFOR
\RETURN [rankedAttList, rankedScoreList]
\end{algorithmic}
\vspace{-1ex}
\end{algorithm}

\subsubsection{Score maximizing greedy conservative mitigation algorithm (DetCons) and its relaxed variant (DetRelaxed)} \label{sec:detconsrelaxed}
While DetGreedy generates rankings with as high score candidates as possible in the ranked list, it may easily fall into an infeasible state (Definition~\ref{def:feasibility}). Hence, it may be desirable to give preference to attribute values that are likely to violate the minimum representation requirement soon enough in the ranking, which is the basis for our next two algorithms. For example, consider a setting with three attribute values and desired proportions of $p_{a_1} = 0.55$, $p_{a_2} = 0.3$, and $p_{a_3} = 0.15$. Suppose that the top 9 results consist of 5 candidates with $a_1$, 3 with $a_2$, and 1 with $a_3$. For $k=10$, the minimum representation requirement is already satisfied for all three attributes while the maximum representation requirements are not met for $a_1$ and $a_3$. However, we can see that the minimum representation requirement will be violated sooner for $a_1$ (at $k=11$, since $\floor{11 \cdot 0.55} = 6$) compared to $a_3$ (at $k=14$, since $\floor{14 \cdot 0.15} = 2$) under the current allocation, and hence it is preferable to choose a candidate with $a_1$ for the position, $k=10$.

Deterministic Conservative (DetCons) algorithm and its relaxed version, Deterministic Relaxed (DetRelaxed), described in Alg.~\ref{alg:detconsrelaxed}, work as follows. As in the case of DetGreedy, if there are any attribute values for which the minimum representation requirement (Eq.~\ref{eq:ranking_objective_2}) is about to be violated, we choose the one with the highest next score among them. Otherwise, among those attribute values that have not yet met their maximum requirements (Eq.~\ref{eq:ranking_objective_1}), we favor one for which the minimum representation requirement is likely to be violated soon enough in the ranking. In DetCons, we choose the attribute value that minimizes $\frac{\ceil{p_{a_i} \cdot k} }{p_{a_i}}$ (i.e., the (fractional) position at which the minimum representation requirement will be violated). In DetRelaxed, we also make use of the integrality constraints, and attempt to include candidates with higher scores. Specifically, we consider all attribute values that minimize $\ceil{\frac{\ceil{p_{a_i} \cdot k}}{p_{a_i}}}$ and choose the one with the highest score for the next candidate.

\begin{algorithm}[t]
\caption{\small{Score Maximizing Greedy Conservative Mitigation Algorithm (DetCons) and its Relaxed variant (DetRelaxed)}}
\label{alg:detconsrelaxed}
\small
\begin{algorithmic}[1]

\STATE {\bf foreach} $a_i ~ \in ~ a$, $counts[a_i] := 0$
\STATE rankedAttList $:= [ ]$; ~ ~ rankedScoreList $:= [ ]$

\FOR {$k \in \{1, \ldots, k_{max} \}$}
	\STATE belowMin $:= \{ a_i: counts[a_i] < \floor{k \cdot p_{a_i}} \}$
	\STATE belowMax $:= \{ a_i: counts[a_i] \geq \floor{k \cdot p_{a_i}} \textrm{ and } counts[a_i] < \ceil{k \cdot p_{a_i}} \}$
	\IF {belowMin $\neq$ $\emptyset$}
		\STATE nextAtt $:=$ argmax$_{a_i \in \textrm{belowMin}} ~ s_{a_i, counts[a_i]}$
	\ELSE
		\IF{\textcolor{red}{\textbf{\emph{DetCons}}}}
			\STATE nextAtt $:=$ argmin$_{a_i \in \textrm{belowMax}} \frac{\ceil{k \cdot p_{a_i}}}{p_{a_i}}$
		\ELSIF{\textcolor{red}{\textbf{\emph{DetRelaxed}}}}
			\STATE nextAttSet $:=$ argmin$_{a_i \in \textrm{belowMax}} \ceil{\frac{\ceil{k \cdot p_{a_i}}}{p_{a_i}}}$ (i.e., the set of all attribute values in belowMax that minimize this term)
			\STATE nextAtt $:=$ argmax$_{a_i \in \textrm{nextAttSet}} ~ s_{a_i, counts[a_i]}$
		\ENDIF
	\ENDIF
	\STATE rankedAttList[k] $:=$ nextAtt
	\STATE rankedScoreList[k] $:=$ $s_{nextAtt, ~ counts[nextAtt]}$
	\STATE counts[nextAtt]$++$
\ENDFOR
\RETURN [rankedAttList, rankedScoreList]
\end{algorithmic}
\vspace{-1ex}
\end{algorithm}

While the above three algorithms are designed towards meeting the conditions given in Eq.~\ref{eq:ranking_objective_1} and Eq.~\ref{eq:ranking_objective_2}, we can show that DetGreedy is not feasible in certain settings. Although we have not been able to prove that DetCons and DetRelaxed are always feasible, our simulation results (\S\ref{sec:results}) suggest that this may indeed be the case.
\begin{theorem}
\label{th:feas1}
The algorithms DetGreedy, DetCons, and DetRelaxed \textbf{are feasible} whenever the number of possible attribute values for the protected attribute is less than 4, i.e., for $|A| \leq 3$. DetGreedy is \textbf{not guaranteed to be feasible} whenever $|A| \geq 4$.
\end{theorem}
Proof is available in the appendix (\S\ref{sec:proof1}). Next, we present a provably feasible algorithm for fairness-aware ranking, which follows a constrained sorting scheme.
\begin{algorithm}[t]
\caption{\small{Feasible Mitigation Algorithm Based on Interval Constrained Sorting (DetConstSort)}}
\label{alg:detconstsort}
\small
\begin{algorithmic}[1]

\STATE {\bf foreach} $a_i ~ \in ~ a$, counts$[a_i] := 0$
\STATE {\bf foreach} $a_i ~ \in ~ a$, minCounts$[a_i] := 0$
\STATE rankedAttList $:= [ ]$; ~ ~ rankedScoreList $:= [ ]$; ~ ~ maxIndices $:= [ ]$
\STATE lastEmpty $:= 0$; ~ ~ k $:= 0$

\WHILE {lastEmpty $\leq k_{max}$}
	\STATE k++
	\STATE {\bf foreach} $a_i ~ \in ~ a$, tempMinCounts$[a_i] := \floor{k \cdot p_{a_i}}$
	\STATE changedMins $:= \{ a_i: \textrm{ minCounts}[a_i] < \textrm{ tempMinCounts}[a_i] \}$

	\IF {changedMins $\neq$ $\emptyset$}
		\STATE ordChangedMins $:=$ {\em sort} changedMins {\em by} $s_{a_i, ~ counts[a_i]}$ {\em descending}
				
		\FOR {$a_i$ $\in$ ordChangedMins (chosen in the sorted order)}
			\STATE rankedAttList[lastEmpty] $:= a_i$
			\STATE rankedScoreList[lastEmpty]  $:= s_{a_i, ~ counts[a_i]}$
			\STATE maxIndices[lastEmpty] $:=$ k
			\STATE start $:=$ lastEmpty

			\WHILE {start > 0 and  maxIndices[start - 1] $\geq$ start and rankedScoreList[start-1] < rankedScoreList[start]}
				\STATE swap(maxIndices[start - 1],  maxIndices[start])
				\STATE swap(rankedAttList[start - 1],  rankedAttList[start])
				\STATE swap(rankedScoreList[start - 1],  rankedScoreList[start])
				\STATE start$- -$
			\ENDWHILE
			
			\STATE counts[$a_i$]++
			\STATE lastEmpty++
		\ENDFOR

		\STATE minCounts $:=$ tempMinCounts
	\ENDIF
\ENDWHILE

\RETURN [rankedAttList, rankedScoreList]
\end{algorithmic}
\vspace{-1ex}
\end{algorithm}

\subsubsection{Feasible mitigation algorithm which employs interval constrained ordering (DetConstSort)}
Deterministic Constrained Sorting (DetConstSort) algorithm (Alg.~\ref{alg:detconstsort}) also aims to enforce the conditions given in Eq.~\ref{eq:ranking_objective_1} and Eq.~\ref{eq:ranking_objective_2}. However, contrary to the three greedy approaches listed previously, DetConstSort waits for multiple indices of recommendation before deciding on the next attribute value to get a candidate from, and may change its previous decisions to improve the score ordering. The algorithm works as follows:
\small
\begin{compact_numbered_enum}
\item Increase a counter value (starting with 0) until at least one attribute value has increased its minimum representation count requirement per Eq.~\ref{eq:ranking_objective_2}. If there is more than one such attribute value, order them according to descending score of their next candidates.
\item Go over the list of ordered attribute values that have increased their minimum requirement, and for each one:
\begin{compact_numbered_enum}
\item Insert the next candidate from the attribute value to the next empty index in the recommendation list.
\item Swap this candidate towards earlier indices in the list until:
\begin{compact_enum}
\item Either the score of the left candidate (candidate in the earlier index) is larger, or,
\item Maximum index of the left candidate will be violated due to the swap (maximum index of a candidate is the maximum index (position) in the ranking where the candidate from this attribute value can be placed at so that we meet the condition of feasibility, i.e., $count_k(a_i) \geq \floor{p_{a_i} \cdot k}$ per Definition~\ref{def:feasibility}.
\end{compact_enum}
\end{compact_numbered_enum}
\end{compact_numbered_enum}
\normalsize
DetConstSort algorithm can be thought of as solving a more general interval constrained sorting problem where we want to maximize the sorting quality, subject to constraints that some elements cannot go beyond a specific index, as long as there is a solution that satisfies the constraints.

\begin{theorem}
\label{th:feas2}
DetConstSort algorithm is feasible per Definition~\ref{def:feasibility}.
\end{theorem}
Proof is available in the appendix (\S\ref{sec:proof2}).

\subsection{Mapping from Fairness Notions to Desired Attribute Distributions} \label{sec:mappingtofairnessnotions}
Our fairness-aware ranking approach aims to achieve representativeness as determined by the desired distribution over a protected attribute (or multiple attributes, by considering the cross-product of possible values). Next, we discuss how our framework can be used to achieve fairness notions such as equal opportunity \cite{hardt_2016} and demographic parity \cite{dwork_2012} through a careful selection of the desired distribution.

\subsubsection{Achieving Equal Opportunity}\label{sec:equalopportunity}
A predictor function $\hat{Y}$ is said to satisfy equal opportunity~\cite{hardt_2016} with respect to a protected attribute $A$ and true outcome $Y$, if the predictor and the protected attribute are independent conditional on the true outcome being 1 (favorable). That is,
\begin{equation}
\label{eq:eqop1}
p(\hat{Y}=1|A=a_1,Y=1)=...=p(\hat{Y}=1|A=a_l,Y=1)~.
\end{equation}
For a search or recommendation task, we can roughly map our framework to the above fairness notion by assuming that the set of candidates that match the criteria (either explicitly specified in the search request or implicitly for the recommendation task) as ``qualified'' for the task. The true outcome being positive ($Y = 1$) corresponds to a candidate matching the search request criteria (or equivalently, being ``qualified'' for the search request), while the prediction being positive ($\hat{Y}=1$) corresponds to a candidate being presented in the top ranked results for the search request. The equal opportunity notion requires that the fraction of qualified candidates that are included in the top ranked results does not depend on the protected attribute, or equivalently that the proportion of candidates belonging to a given value of the attribute does not vary between the set of qualified candidates and the set of top ranked results. In our framework, this requirement can be met by selecting the desired distribution to be the distribution of the {\em qualified} candidates over the protected attribute. Further, since the top ranked results are chosen from the set of qualified candidates (that is, $\hat{Y}=0$ whenever $Y = 0$), the above choice of the desired distribution can also be viewed as meeting the requirement of equalized odds~\cite{hardt_2016}.

\subsubsection{Achieving Demographic Parity}
Demographic parity (or statistical parity)~\cite{dwork_2012} requires that the predictor function $\hat{Y}$ be independent of the protected attribute $A$, that is,
\[
p(\hat{Y}=1|A=a_1)=...=p(\hat{Y}=1|A=a_l)\textrm{, and,}
\]
\begin{equation}
\label{eq:dempar1}
p(\hat{Y}=0|A=a_1)=...=p(\hat{Y}=0|A=a_l)~.
\end{equation}
In our framework, we can show that this requirement can be met by selecting the desired distribution to be the distribution of {\em all} candidates over the protected attribute (following a similar argument as in \S\ref{sec:equalopportunity}). Demographic parity is an important consideration in certain application settings, although it does not take qualifications into account and is known to have limitations (see \cite{dwork_2012, hardt_2016}). For example, in the case of gender, demographic parity would require that the top results always reflect the gender distribution over all candidates, irrespective of the specific search or recommendation task.

%% file: results.tex
\section{Evaluation and Deployment in Practice} \label{sec:results}
In this section, we evaluate our proposed fairness-aware ranking framework via both offline simulations, and through our online deployment in \emph{LinkedIn Recruiter} application.

\subsection{Simulation Results} \label{sec:resultssub}
Next, we present the results of evaluating our proposed fairness-aware re-ranking algorithms via extensive simulations. Rather than utilizing a real-world dataset, we chose to use simulations for the following reasons:
\begin{compact_numbered_enum}
\item \textbf{To be able to study settings where there could be several possible values for the protected attribute.} Our simulation framework allowed us to evaluate the algorithms over attributes with up to 10 values (e.g., <gender, age group> which could assume 9 values with three gender values (male, female, and other/unknown) and three age groups), and also study the effect of varying the number of possible attribute values. In addition, we generated many randomized settings covering a much larger space of potential ranking situations, and thereby evaluated the algorithms more comprehensively.
\item \textbf{Evaluating the effect of re-ranking on a utility measure in a dataset collected from the logs of a specific application is often challenging due to \emph{position bias} \cite{joachims_2005}.} Utilizing a simulation framework allows random assignment of relevance scores to the ranked candidates (to simulate the scores of a machine learned model) and directly measure the effect of fairness-aware re-ranking as compared to score based ranking.
\end{compact_numbered_enum}

{\noindent \bf Simulation framework}:
\small
\begin{compact_numbered_enum}
\item For each possible number of attribute values ($2 \leq |A| \leq 10$):
\begin{compact_numbered_enum}
\item Generate a set $P$ of 100K random categorical probability distributions of size $|A|$ each. Each probability distribution $P_j \in P$ is generated by choosing $|A|$ i.i.d. samples from the uniform distribution over $(0,1)$ and normalizing the sum to equal 1. Each $P_j$ represents a possible desired distribution over the set $A$ of attribute values.
\item For each $P_j \in P$:
\begin{compact_numbered_enum}
\item For each attribute value in $A$, generate 100 random candidates whose scores are chosen i.i.d. from the uniform distribution over $(0,1)$, and order them by decreasing scores. We replicate this step 10 times (resulting in 1M distinct ranking tasks for each choice of $|A|$).
\item Run each proposed fairness-aware ranking algorithm to get a fairness-aware re-ranked list of size 100, with the desired distribution $P_j$ and the generated random candidate lists for each attribute value as inputs.
\end{compact_numbered_enum}
\end{compact_numbered_enum}
\end{compact_numbered_enum}
\normalsize
For each ranking task generated by the above framework, we compute the proposed bias measures such as InfeasibleIndex  (Eq.~\ref{eq:infeasible_index}), MinSkew (Eq.~\ref{eq:minskew_at_k}), and NDKL (Eq.~\ref{eq:ndkl}), as well as Normalized Discounted Cumulative Gain\footnote{~ NDCG is defined over a ranked list of candidates $\taur$ as follows:\\
$\textrm{NDCG}(\taur) = \frac{1}{Z} \times \sum_{i=1}^{|\taur|} \frac{u(\taur[i])}{\log(i + 1)}$,
where $u(\taur[i])$ is the relevance for the candidate in $i^{th}$ position of $\taur$. In our simulations, we treat the score of each candidate as the relevance, whereas in real-world applications, relevance could be obtained based on human judgment labels or user response (e.g., whether or the extent to which the user liked the candidate).
$Z$ is the normalizing factor corresponding to the discounted cumulative gain for the best possible ranking $\taur^*$ of candidates, i.e., $Z = \sum_{i=1}^{|\taur^*|} \frac{u(\taur^*[i])}{\log(i + 1)}$.
}
(NDCG) \cite{jarvelin_2002} as a measure of the ``ranking utility'' where we treat the scores of candidates as their relevance. We report the results in terms of the average computed over all ranking tasks for a given choice of the number of attribute values.\\

\begin{figure}[ht]
\centering
\includegraphics[width=2.85in]{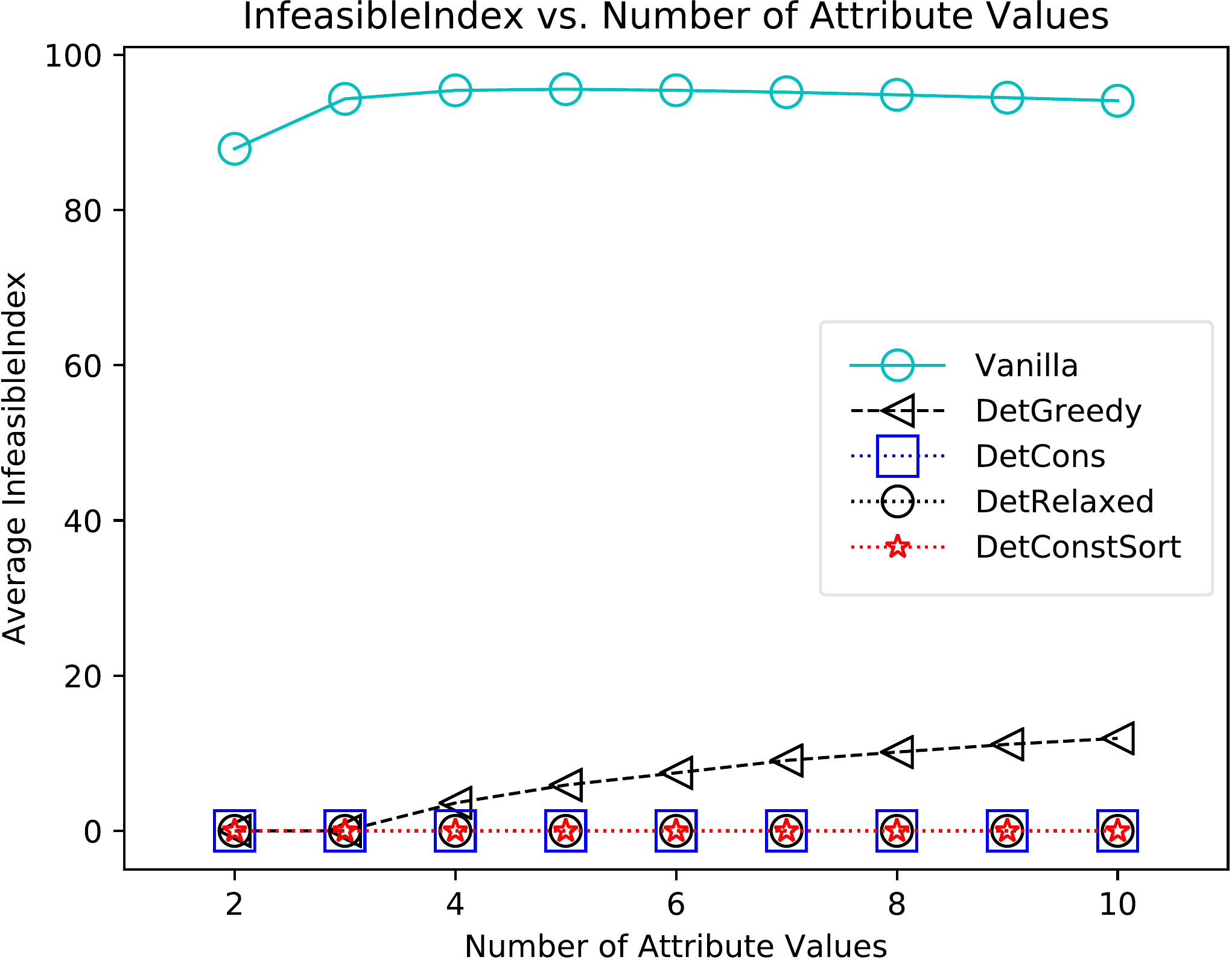}
\vspace{-10pt}
\caption{InfeasibleIndex Measure Results}
\label{fig:infIndex_All}
\end{figure}

\begin{figure}[ht]
\centering
\includegraphics[width=2.85in]{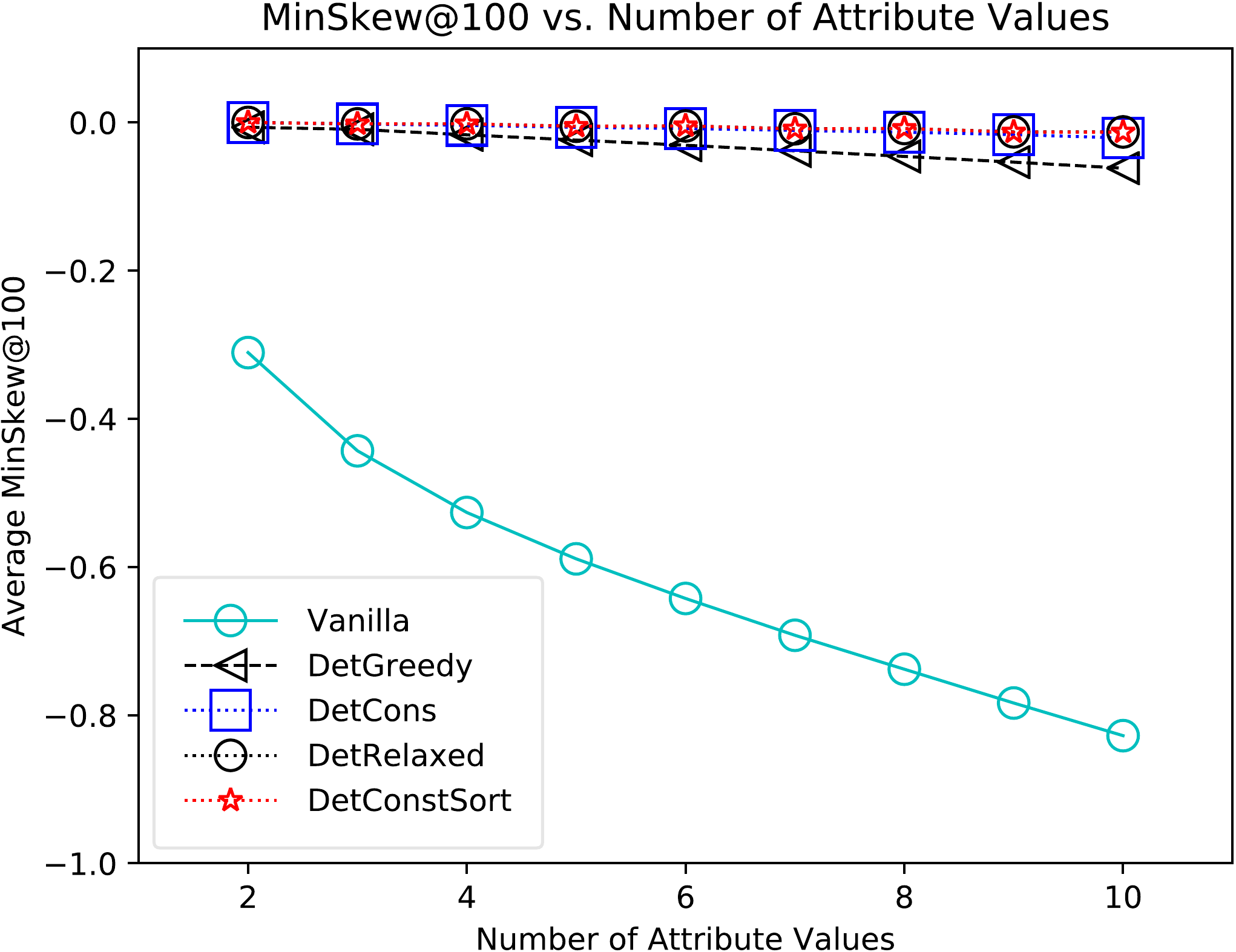}
\vspace{-10pt}
\caption{MinSkew@100 Measure Results}
\label{fig:minSkew_All}
\end{figure}

\begin{figure}[ht]
\vspace{-5pt}
\centering
\includegraphics[width=2.85in]{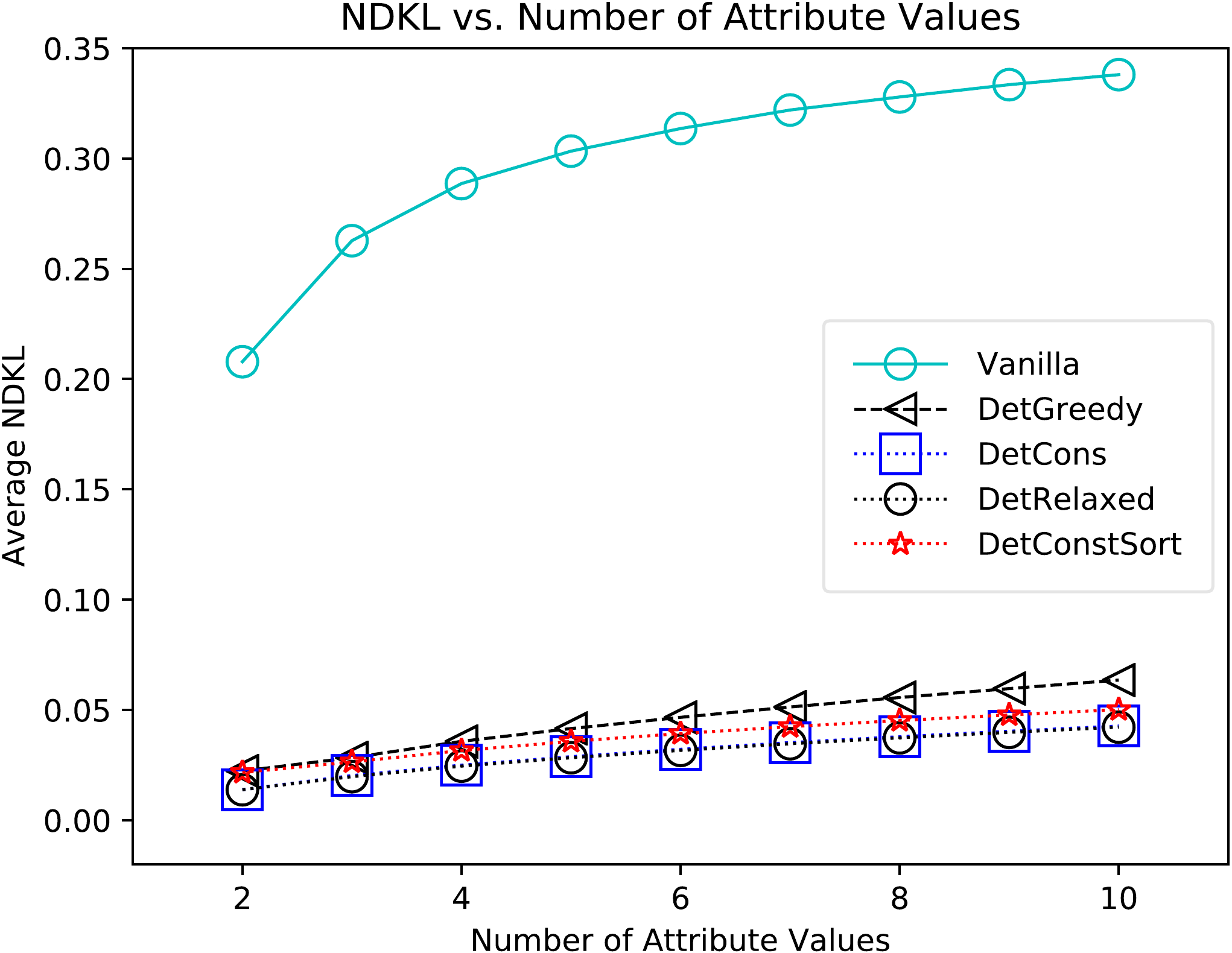}
\vspace{-10pt}
\caption{NDKL Measure Results}
\label{fig:ndkl_All}
\end{figure}

\begin{figure}[ht]
\vspace{-5pt}
\centering
\includegraphics[width=2.85in]{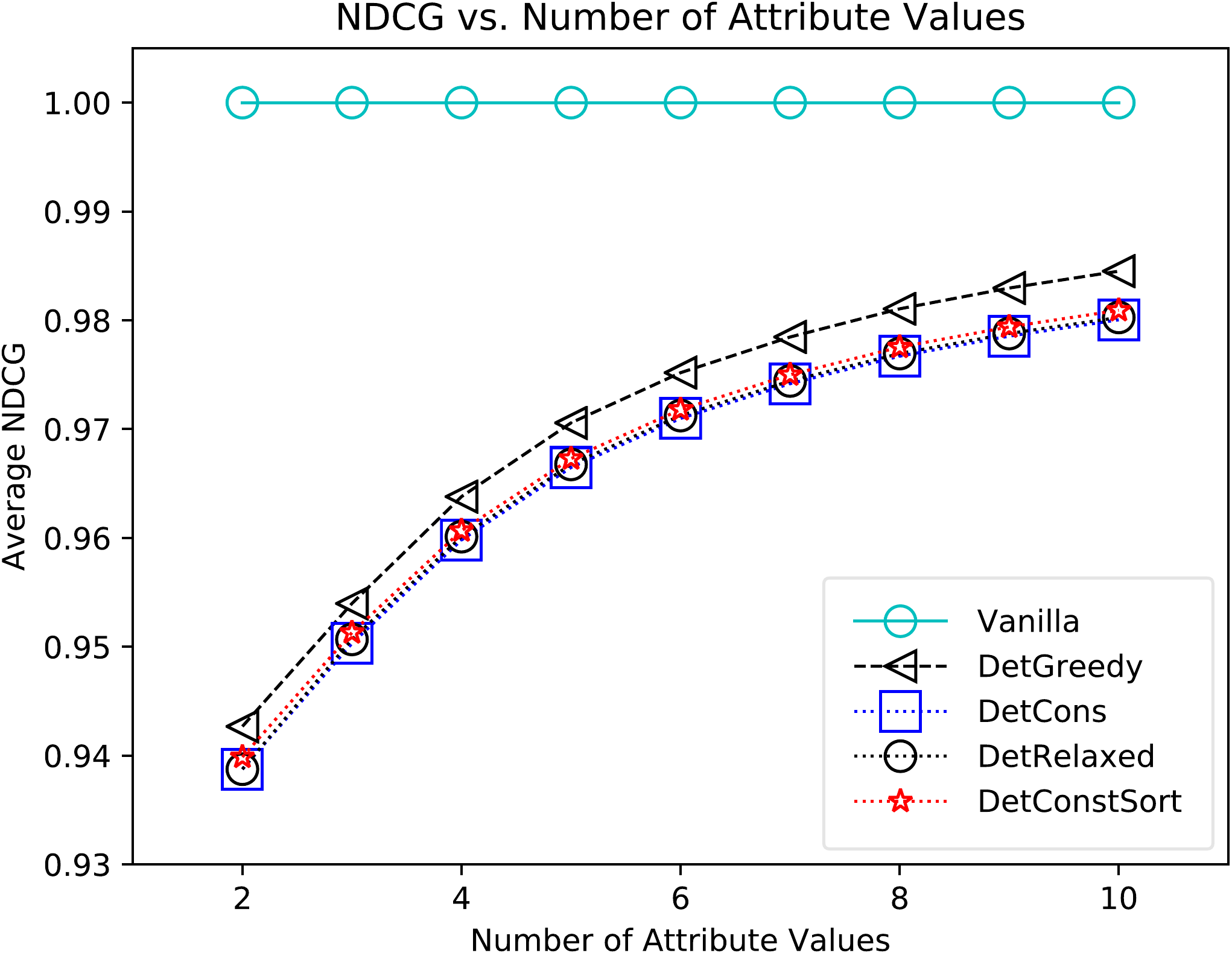}
\vspace{-10pt}
\caption{NDCG@100 Measure Results}
\label{fig:ndcg_All}
\end{figure}

{\noindent \bf Results}: Figures~\ref{fig:infIndex_All} through \ref{fig:ndcg_All} give the bias and utility results as a function of the number of attribute values for the proposed algorithms per the simulation framework.

From Figure \ref{fig:infIndex_All}, we can see that all our proposed algorithms are feasible for attributes with up to 3 possible values (which is in confirmation with our feasibility results (\S\ref{sec:algorithms})). We observed similar results for InfeasibleCount measure (Eq.~\ref{eq:infeasible_count}; results given in \S\ref{sec:moreresults}). We observe that DetConstSort is also feasible for all values of $|A|$ (in agreement with the theorem in \S\ref{sec:algorithms}). Furthermore, for DetGreedy, InfeasibleIndex measure increases with the number of possible attribute values, since it becomes harder to satisfy Eq.~\ref{eq:ranking_objective_2} for a large number of attribute values. We can also see that both DetCons and DetRelaxed are feasible for all values of $|A|$, which, although not proven, gives strong evidence to their general feasibility.

Figure \ref{fig:minSkew_All} presents the results for MinSkew@100 measure. We observed similar results for MaxSkew measure (Eq.~\ref{eq:maxskew_at_k}; results given in \S\ref{sec:moreresults}). DetCons, DetRelaxed, and DetConstSort algorithms perform quite similarly, and overall better than DetGreedy, as expected. All the fairness-aware algorithms perform much better compared to the baseline score-based (vanilla) ranking.

The results for NDKL measure, presented in Figure~\ref{fig:ndkl_All}, show that the look-ahead algorithms, DetCons and DetRelaxed, perform slightly better than DetConstSort.

For utility evaluation, we computed the NDCG@100 of the generated rankings to see whether re-ranking causes a large deviation from a ranking strategy based fully on the relevance scores. Figure~\ref{fig:ndcg_All} shows that DetGreedy performs significantly better compared to the rest of fairness-aware ranking algorithms in terms of utility. DetConstSort also performs slightly better compared to the look-ahead algorithms (DetCons and DetRelaxed). Note that the vanilla algorithm ranks purely based on scores, and hence has a constant NDCG of 1.

\begin{figure*}[ht]
\centering
\includegraphics[width=7.0in]{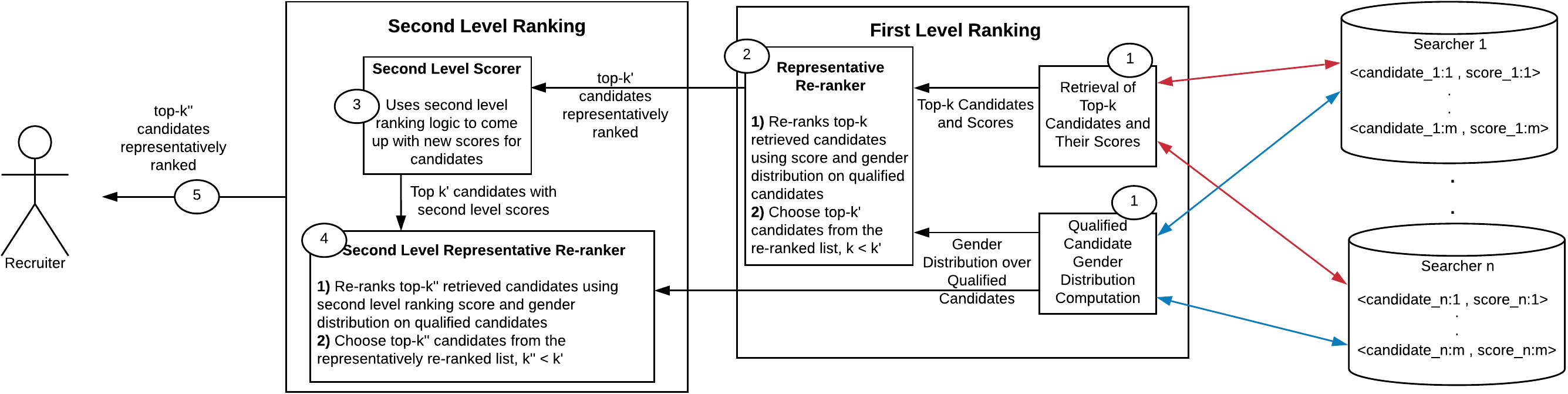}
\vspace{-10pt}
\caption{Online Architecture for Gender-Representative Ranking at LinkedIn}
\label{fig:online_architecture}
\end{figure*}

Overall, DetGreedy has very competitive performance in terms of fairness measures and generates ranked lists with the highest utility. However, if the requirements of minimum representation for each attribute value are strict, we would be confined to DetCons, DetRelaxed, and DetConstSort (which happens to be the only algorithm we have theoretically proven to be feasible). Among those algorithms that did generate consistently feasible rankings in our simulations, DetConstSort performed slightly better in terms of utility. In terms of fairness measures though, we did not observe considerable difference amongst DetCons, DetRelaxed, and DetConstSort. In summary, there is no single ``best'' algorithm, and hence it would be desirable to carefully study the fairness vs. utility trade-offs in the application setting (e.g., by  performing A/B testing) and thereby select the most suitable of these algorithms.

\subsection{Online A/B Testing Results and Deployment in LinkedIn Talent Search}\label{sec:deployment}
We have implemented the proposed framework as part of the \emph{LinkedIn Recruiter} product to ensure gender-representative ranking of candidates. This product enables recruiters and hiring managers to source suitable talent for their needs, by allowing them to perform search and reach out to suitable candidates. Given a search request, this system first retrieves the candidates that match the request out of a pool of hundreds of millions of candidates, and then returns a ranked list of candidates using machine-learned models in multiple passes (see Figure~\ref{fig:online_architecture}, explained in \S\ref{sec:arch}). For each search request, the desired gender distribution over the ranked candidate list is chosen to be the gender distribution over the set of candidates that meet (i.e., qualify for) the search criteria provided by the user of the system (recruiter). The candidate set retrieval and scoring, as well as the computation of the desired distribution, is performed in a distributed manner using LinkedIn's {\em Galene} search engine~\cite{Sankar14}. Computing the desired distribution in this manner can be thought of as corresponding to achieving \emph{equality of opportunity} per discussion in \S\ref{sec:mappingtofairnessnotions}. We utilized Algorithm~\ref{alg:detgreedy} (\emph{DetGreedy}) in our online deployment due to its implementation simplicity and practicality considerations with A/B testing multiple algorithms (such as ensuring sufficient statistical power). Also, we observed in \S\ref{sec:resultssub} that it provided the highest utility and good performance in terms of fairness measures, especially for protected attributes with low cardinality like gender (per Theorem~\ref{th:feas1}, DetGreedy is feasible for attributes with up to three values, and gender fits this description).

The results of the A/B test which we performed over three weeks within 2018 with hundreds of thousands of Recruiter users is presented in Table~\ref{tab:abresults}. In this experiment, a randomly chosen 50\% of Recruiter users were presented with results from the fairness-aware ranking approach while the rest were presented with the vanilla ranking of candidates based on the scores from the ML model, which is optimized for the likelihood of making a successful hire. Please refer \cite{geyik_2018} and the references therein for a detailed description of ML models used in LinkedIn Talent Search. Our fair re-ranking approach has ensured that more than {\bf 95\% of all the searches are representative} of any gender compared to the qualified population of the search (i.e., the ranking is {\em feasible} per Definition~\ref{def:feasibility} in 95\% of the cases), which is nearly a {\bf 3X improvement}. Furthermore, \emph{MinSkew} (Skew for the most disadvantaged gender group within the results of a search query) over top 100 candidates, averaged over all search requests, is now very close to 0 (we achieved similar results over top 25, top 50, etc., and for other fairness measures). In other words, ranked lists of candidates presented are representative in practice. We did not observe any statistically significant change in business metrics, such as the number of inMails sent [messages from recruiters to candidates] or inMails accepted [messages from recruiters to candidates, answered back with positive responses] (only relative values are presented in Table~\ref{tab:abresults}), meaning that ensuring representation did not negatively impact the customers' success metrics or quality of the presented candidates for our application. Based on these results, we decided to ramp the re-ranking approach to 100\% of Recruiter users worldwide. We direct the interested reader to our engineering blog post~\cite{reprank_blogpost} for further details.

\begin{table}[!ht]
\caption{Online A/B Test Results}
\vspace{-10pt}
\small
\begin{tabular}{c|c|c}
\hline\hline
\textbf{Metric} & \textbf{Vanilla} & \textbf{Fairness-aware} \\ \hline\hline
Queries with representative results & {\bf 33\%} & {\bf 95\%} \\ \hline
Average MinSkew@100 & -0.259 & -0.011 (p-value < 1e-16) \\ \hline
InMails Sent & - & $\pm$ 1\% (p-value $>$ 0.5) \\ \hline
InMails Accepted & - & $\pm$ 1\% (p-value $>$ 0.5) \\ \hline
\hline
\end{tabular}
\label{tab:abresults}
\end{table}

\subsection{Lessons Learned in Practice}\label{sec:lessons}
{\noindent \bf Post-Processing vs. Alternate Approaches}: Broadly, there are three technical approaches for mitigating algorithmic bias in machine learning systems:
\begin{compact_enum}
\item \emph{Pre-processing} includes the efforts prior to model training such as representative training data collection and modifying features or labels in the training data (e.g. \cite{calmon_2017}).
\item \emph{Modifying the training process} to generate a bias-free model (e.g., \cite{asudeh_2017}).
\item \emph{Post-processing} includes the modification of the results of a trained machine learning model, using techniques such as calibration of regression or classification output and re-ranking of results (e.g., \cite{zehlike_2017}).
\end{compact_enum}

We decided to focus on post-processing algorithms due to the following practical considerations which we learned over the course of our investigations. First, applying such a methodology is agnostic to the specifics of each model and therefore scalable across different model choices for the same application and also across other similar applications. Second, in many practical internet applications, domain-specific business logic is typically applied prior to displaying the results from the ML model to the end user (e.g., prune candidates working at the same company as the recruiter), and hence it is more effective to incorporate bias mitigation as the very last step of the pipeline. Third, this approach is easier to incorporate as part of existing systems, as compared to modifying the training algorithm or the features, since we can build a stand-alone service or component for post-processing without significant modifications to the existing components. In fact, our experience in practice suggests that post-processing is easier than eliminating bias from training data or during model training (especially due to redundant encoding of protected attributes and the likelihood of both the model choices and features evolving over time). However, we remark that efforts to eliminate/reduce bias from training data or during model training can still be explored, and can be thought of as complementary to our approach, which functions as a ``fail-safe''.\\

{\noindent \bf Socio-technical Dimensions of Bias and Fairness}: Although our fairness-aware ranking algorithms are agnostic to how the desired distribution for the protected attribute(s) is chosen and treat this distribution as an input, the choice of the desired bias / fairness notions (and hence the above distribution) needs to be guided by ethical, social, and legal dimensions. As discussed in \S\ref{sec:mappingtofairnessnotions}, our framework can be used to achieve different fairness notions depending on the choice of the desired distribution. Guided by LinkedIn's goal of creating economic opportunity for every member of the global workforce and by a keen interest from LinkedIn's customers in making sure that they are able to source diverse talent, we adopted a ``diversity by design'' approach for LinkedIn Talent Search, and took the position that the top search results for each query should be representative of the broader qualified candidate set~\cite{reprank_blogpost}. The representative ranking requirement is not only simple to explain (as compared to, say, approaches based on statistical significance testing (e.g.,~\cite{zehlike_2017})), but also has the benefit of providing consistent experience for a recruiter or a hiring manager, who could learn about the gender diversity of a certain talent pool (e.g., sales associates in Anchorage, Alaska) and then see the same distribution in the top search results for the corresponding search query. Our experience also suggests that building consensus and achieving collaboration across key stakeholders (such as product, legal, PR, engineering, and AI/ML teams) is a prerequisite for successful adoption of fairness-aware approaches in practice~\cite{fairnessTutorialWWW19}.

%% file: relatedwork.tex
\section{Related Work} \label{sec:background}

\input{relatedwork_subsec_1}

\input{relatedwork_subsec_2}

%% file: relatedwork_subsec_1.tex
There has been an extensive study of algorithmic bias and discrimination across disciplines such as law, policy, and computer science (e.g., see~\cite{hajian_2016_tutorial, zliobaite2017measuring, friedman_1996} and the references therein). Many recent studies have investigated two different notions of fairness: (1) {\em individual fairness}, which requires that similar people be treated similarly~\cite{dwork_2012}, and (2) {\em group fairness}, which requires that the disadvantaged group be treated similarly to the advantaged group or the entire population~\cite{pedreschi_2008,pedreschi_2009}. While some studies focus on identifying and quantifying the extent of discrimination (e.g.,~\cite{angwin_2016, caliskan_2017, pedreschi_2008}), others study mitigation approaches in the form of fairness-aware algorithms (e.g.,~\cite{calders_2010, celis_2016, celis_2017, corbett_2017, dwork_2012, friedler_2016, friedler2018comparative, hajian_2013, hajian_2014, hardt_2016, jabbari_2017, kamiran_2010, kleinberg_2017, woodworth2017learning, zafar2017fairness, zehlike_2017, zemel_2013}) and inherent trade-offs and limitations in achieving different notions of fairness and non-discrimination~\cite{corbett_2017, dwork_2012, friedler_2016, kleinberg_2017}. Formal definitions of group fairness include demographic parity \cite{dwork_2012} and equal opportunity \cite{hardt_2016}. As discussed in \S\ref{sec:mappingtofairnessnotions}, our framework supports these two definitions through appropriate choice of the desired distribution. We remark that there is extensive work in social science, philosophy, and legal literature on discrimination and fairness. We defer the reader to \cite{barocas2016big} for a discussion from a legal perspective and \cite{arneson_2018} for a discussion of four different frameworks of equal opportunity.

%% file: relatedwork_subsec_2.tex
Our work is closely related to recent literature on fairness in ranking~\cite{celis_2017, yang_2017, zehlike_2017, singh_2018, biega_2018, asudeh_2017, castillo2019fairness}. A method to assist the algorithm designer to generate a fair linear ranking model has been proposed in~\cite{asudeh_2017}. With respect to a fixed set of items, given a weight vector for ranking, the method in~\cite{asudeh_2017} computes a similar vector that meets fairness requirements. This approach is not applicable in our setting since it assumes that the candidate set of items to be ranked is fixed, whereas this set depends on the query in our case. Further, since it is limited to linear models and requires modifying the weight vector, this approach would be hard to scale across different model choices in practice (see \S\ref{sec:lessons}). The problem of achieving individual equity-of-attention fairness in rankings, along with a mechanism for achieving amortized individual fairness, has been proposed in~\cite{biega_2018}. While this work aims to achieve individual fairness amortized across many rankings, our focus is on ensuring that each ranking meets the group fairness requirements specified using a desired distribution. Our proposed algorithms are designed to mitigate biases in the ranked results for each query by achieving a desired distribution over a protected attribute. Algorithms for ranking in the presence of fairness constraints, specified as the maximum (or minimum) number of elements of each class that can appear at any position in the ranking, have been proposed in~\cite{celis_2017, zehlike_2017}. Zehlike et al.~\cite{zehlike_2017} propose a fair top-k ranking algorithm focusing on a required representation proportion for a {\em single} under-represented group. Our proposed algorithms allow attributes with many possible values as opposed to just binary attributes, hence constituting a more general framework, and can handle representation constraints corresponding to an arbitrary, desired categorical distribution. FA*IR algorithm proposed in~\cite{zehlike_2017} can be thought of as similar to our DetGreedy method that works only for a binary protected attribute, with a considerable difference in the minimum representation requirement computation. Celis et al.~\cite{celis_2017} present a theoretical investigation of fair ranking computation with constraints on the maximum and the minimum number of elements possessing an attribute value that can be present at each rank. In contrast, we provide relatively easy-to-explain and easy-to-implement algorithms since our work is motivated by the desire to implement and deploy in practice (see \S\ref{sec:deployment} and \S\ref{sec:lessons}). Further, by presenting an empirical evaluation of the trade-off between fairness and business metrics, we enable practitioners to select a suitable algorithm for their application needs. A framework for formulating fairness constraints on rankings, and an associated probabilistic algorithm for computing utility maximizing fair ranking have been proposed in~\cite{singh_2018}. This method requires solving a linear program with a large number of variables and constraints ($N^2$ where $N$ denotes the number of candidates to be ranked), and hence does not seem feasible in a practical search/recommendation system with strict latency requirements. Fairness measures for ranking have been proposed in~\cite{yang_2017}, which we have extended for our setting (\S\ref{sec:evaluatingBias}). Finally, ours is the first large-scale deployed framework for ensuring fairness in ranked results.

%% file: conclusion.tex
\section{Conclusion}\label{sec:conclusion}
Motivated by the desire for creating fair opportunity for all users being ranked as part of search and recommendation systems and the consequent need for measuring and mitigating algorithmic bias in the underlying ranking mechanisms, we proposed a framework for fair re-ranking of results based on desired proportions over one or more protected attributes. We proposed several measures to quantify bias with respect to protected attributes such as gender and age, and presented fairness-aware ranking algorithms. We demonstrated the efficacy of these algorithms in reducing bias without affecting utility, and compared their performance via extensive simulations. We also deployed the proposed framework for achieving gender-representative ranking in LinkedIn Talent Search, where our approach resulted in huge improvement in the fairness metrics (nearly 3X increase in the number of queries with representative results) without impacting the business metrics. In addition to being the first web-scale deployed framework for ensuring fairness in the hiring domain, our work contains insights and lessons learned in practice that could be of interest for researchers and practitioners working on various web-scale search and recommendation systems.

A potential direction for future work is to further study fairness and utility guarantees of the proposed algorithms, as well as to extend with other\cut{ deterministic and non-deterministic} algorithmic variants. While we have experimented with synthetic datasets to capture a wide range of parameter choices compared to what a real-world dataset might include, experiments with real datasets could be performed to complement our empirical study. Our approach assumes that the protected attribute values of candidates as well as the desired proportions are provided as part of the input; these assumptions may not hold in certain application settings. A fruitful direction is to understand the social dimension of how to reliably obtain the protected attribute values of users, if not readily available, and how to design the fairness-aware ranking algorithms in an incentive-compatible fashion in case these values are self-reported. A related direction is to study the social question of how the desired proportions (or fairness notions) should be chosen for different classes of practical applications.

%% file: acknowledgments.tex
We would like to thank other members of LinkedIn Careers and Talent Solutions teams for their collaboration while deploying our system in production, and in particular, Patrick Driscoll, Gurwinder Gulati, Rachel Kumar, Divyakumar Menghani, Chenhui Zhai, and Yani Zhang for working closely with us during the development of the product. We also thank Deepak Agarwal, Erik Buchanan, Patrick Cheung, Gil Cottle, Nadia Fawaz, Joshua Hartman, Heloise Logan, Lei Ni, Ram Swaminathan, Ketan Thakkar, Janardhanan Vembunarayanan, Hinkmond Wong, Lin Yang, and Liang Zhang for insightful feedback and discussions.

%% file: appendix.tex
\section{Appendix}\label{sec:appendix}

\subsection{Results for InfeasibleCount and MaxSkew Measures} \label{sec:moreresults}
\begin{figure}[ht]
\centering
\includegraphics[width=2.85in]{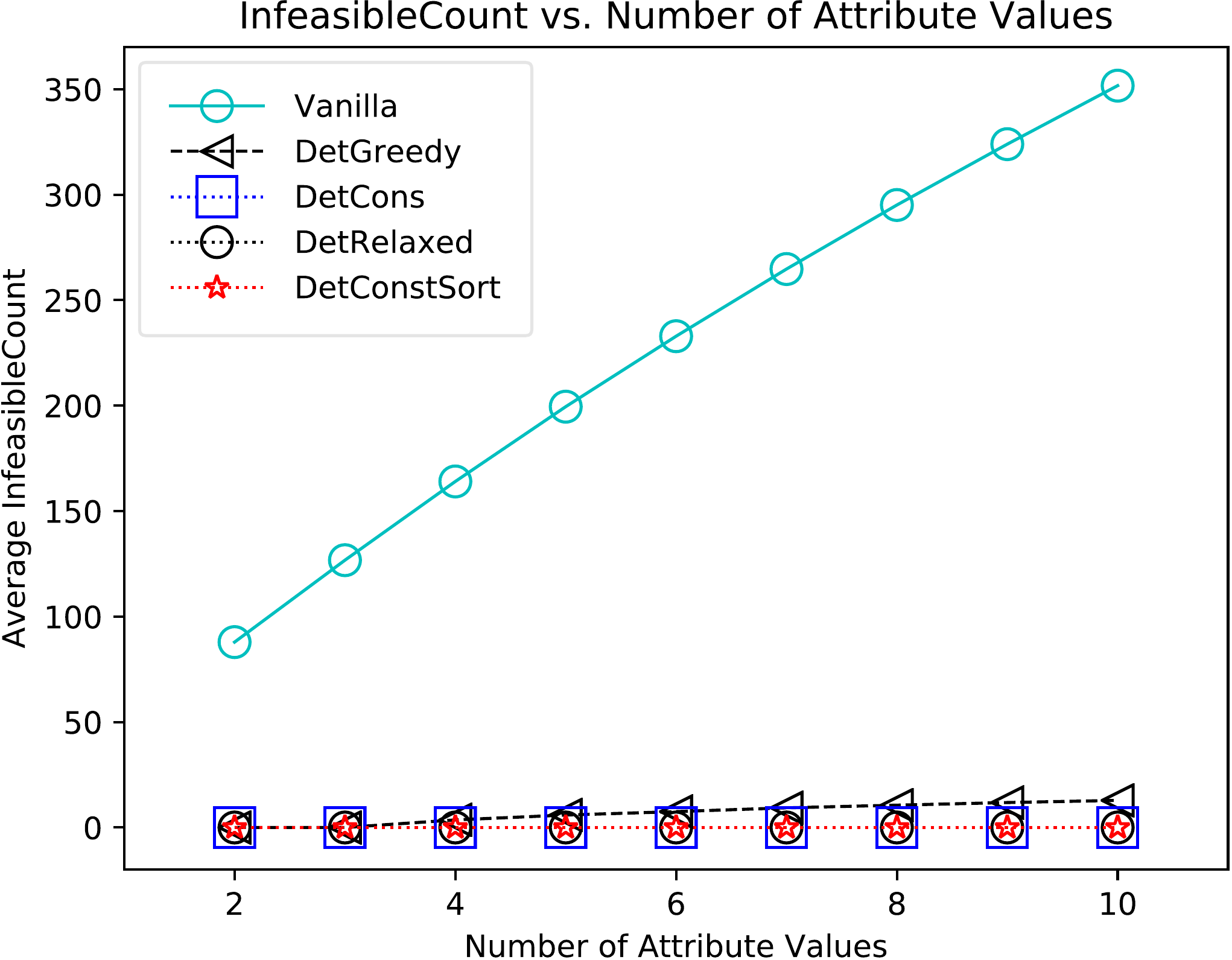}
\vspace{-10pt}
\caption{InfeasibleCount Measure Results}
\label{fig:infCount_All}
\end{figure}

\begin{figure}[ht]
\vspace{-5pt}
\centering
\includegraphics[width=2.85in]{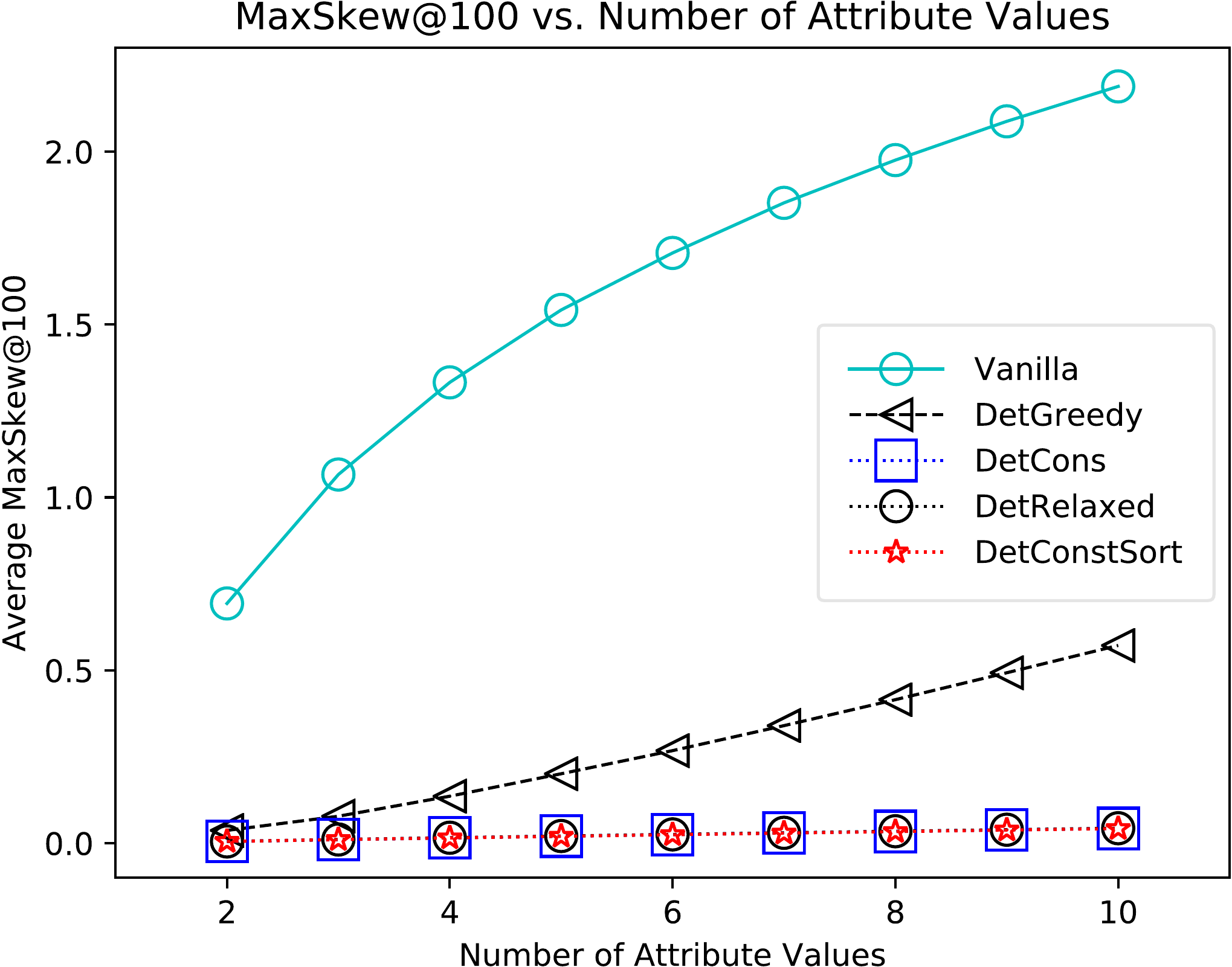}
\vspace{-10pt}
\caption{MaxSkew@100 Measure Results}
\label{fig:maxSkew_All}
\end{figure}

For the continuation of \S\ref{sec:resultssub}, we present the results for InfeasibleCount (Eq.~\ref{eq:infeasible_count}) and MaxSkew@100 (Eq.~\ref{eq:maxskew_at_k}) in Figures \ref{fig:infCount_All} and \ref{fig:maxSkew_All}.

\subsection{Proof of Theorem \ref{th:feas1}} \label{sec:proof1}
\emph{The algorithms DetGreedy, DetCons, and DetRelaxed \textbf{are feasible} whenever the number of possible attribute values for the protected attribute is less than 4, i.e., for $|A| \leq 3$. DetGreedy is \textbf{not guaranteed to be feasible} whenever $|A| \geq 4$}.
\begin{proof}
First, we prove that all three algorithms are feasible for $|A| \leq 3$. Since $|A| = 1$ corresponds to the trivial case of all candidates possessing the same attribute value, we focus on $|A| \in \{2,3\}$. Further, we assume that $0 < p_{a_i} < 1 ~ \forall ~ a_i$, since (1) any attribute value with $p_{a_i} = 0$ does not affect feasibility and would be ignored by our algorithms, and (2) if $p_{a_i} = 1$ for some $a_i$, feasibility would be trivially satisfied since our algorithms would only include candidates possessing $a_i$.

We provide a proof by contradiction. Suppose that there exists $k \geq 0$ such that the ranking was feasible till position $k$, but became infeasible when deciding the attribute value to be chosen for position $k+1$. Note that the ranking is always feasible at the first position ($k=1$) since for any $a_i$, the minimum count requirement is $\floor{p_{a_i} \cdot 1} = 0$. It follows that there are at least two attribute values, say $a_1$ and $a_2$ (without loss of generality), for which the minimum count requirement is about to be violated at $k+1$. In other words, the candidate for position $k+1$ needs to possess both $a_1$ and $a_2$, and since this is impossible, the ranking would become infeasible. Further, 
\begin{align}
\label{eq:a_2_proof_1}
\ceil{p_{a_1} \cdot k} = \floor{p_{a_1} \cdot (k+1)} > count_{k}(a_1) \geq \floor{p_{a_1} \cdot k} ~ and,\\
\label{eq:a_2_proof_2}
\ceil{p_{a_2} \cdot k} = \floor{p_{a_2} \cdot (k+1)} > count_{k}(a_2) \geq \floor{p_{a_2} \cdot k}.
\end{align}
Note that $p_{a_1} \cdot k$ cannot be an integer, as otherwise $\floor{p_{a_1} \cdot (k+1)} = p_{a_1} \cdot k + \floor{p_{a_1}} = p_{a_1} \cdot k$ (using $p_{a_1} < 1$), which goes against our assumption that the ranking became infeasible at $k+1$. By a similar argument, $p_{a_2} \cdot k$ also cannot be an integer. Hence, we have: $\ceil{p_{a_1} \cdot k} = \floor{p_{a_1} \cdot k} + 1 = \floor{p_{a_1} \cdot (k+1)}$ (similarly for $a_2$). Consequently, we also have: $count_{k}(a_1) = \floor{p_{a_1} \cdot k}$ and $count_{k}(a_2) = \floor{p_{a_2} \cdot k}$\\

{\noindent \bf Case 1: $|A| = 2$}: Since the number of candidates included till position $k$ equals $k$, we have: $k = count_{k}(a_1) + count_{k}(a_2) =  \floor{p_{a_1} \cdot k} + \floor{p_{a_2} \cdot k} < {p_{a_1} \cdot k} + {p_{a_2} \cdot k} = (p_{a_1}  + p_{a_2}) \cdot k = 1 \cdot k = k$, which is a contradiction.

{\noindent \bf Case 2: $|A| = 3$}: Since the number of candidates included till position $k$ equals $k$, and since $count_{k}(a_1) = \floor{p_{a_1} \cdot k}$ and $count_{k}(a_2) = \floor{p_{a_2} \cdot k}$, it follows that $count_{k}(a_3) = \ceil{p_{a_3} \cdot k}$. This is because $\floor{p_{a_1} \cdot k} + \floor{p_{a_2} \cdot k} + \floor{p_{a_3} \cdot k} < k$ (recall that $p_{a_1} \cdot k$ and $p_{a_2} \cdot k$ cannot be integers), and our algorithms do not allow $count_{k}(a_3)$ to exceed $\ceil{p_{a_3} \cdot k}$. Therefore, 
\[
(p_{a_1} \cdot k - \floor{p_{a_1} \cdot k}) + (p_{a_2} \cdot k - \floor{p_{a_2} \cdot k}) = ~~~~~~~~~~~~~~~~~~~~~~~~~~~~~~~~~~~~~~~~~~~~~~~~~~
\]
\[
~~~~~~~~~~~~~~~~~~~~~~~~~~~~~~~~~~~~~~~~~~~~~~~~~~ \ceil{p_{a_3} \cdot k} - p_{a_3} \cdot k < 1 ~, \textrm{and hence}~,
\]
\[
p_{a_1} + (p_{a_1} \cdot k - \floor{p_{a_1} \cdot k}) + p_{a_2} + (p_{a_2} \cdot k - \floor{p_{a_2} \cdot k}) < 1 + (p_{a_1} + p_{a_2}).
\]
However, we know that $p_{a_1} + (p_{a_1} \cdot k) = p_{a_1} \cdot (k+1) \geq 1+ \floor{p_{a_1} \cdot k}$ (per conditions in \ref{eq:a_2_proof_1} and \ref{eq:a_2_proof_2}). Hence,
\[
2 \leq p_{a_1} + (p_{a_1} \cdot k - \floor{p_{a_1} \cdot k}) + p_{a_2} + (p_{a_2} \cdot k - \floor{p_{a_2} \cdot k}) ~~~~~~~~~~~~~~~~~~~~
\]
\[
~~~~~~~~~~~~~~~~~~~~~~~~~~~~~~ < 1 + (p_{a_1} + p_{a_2}) ~ \textrm{, so that,}
\]
\[
1 < p_{a_1} + p_{a_2} ~, ~~~~~~~~~~~~~~~~~~~~~~~~~~~~~~~~~~~~~~~~
\]
which is a contradiction, since we know that $\sum p_{a_i} = 1$.

Thus, it follows that the three algorithms are always feasible for $|A| \leq 3$.

We show that the DetGreedy algorithm is not feasible for $|A| \geq 4$ via a simple counter-example. In Table~\ref{table:infeasibility_detgreedy_example}, we list four attribute values with their corresponding desired proportions, and one candidate from each of them (we only give the scores in the brackets). The algorithm chooses $a_4$ and $a_3$ for the first two positions in the ranking. However, for the third position, both $a_1$ and $a_2$ require a candidate from them to be chosen, which makes the ranked list infeasible. Since for any choice of $|A| > 4$, we can extend the above counter-example (by setting $p_{a_i} = 0$ for $i > 4$), it follows that DetGreedy algorithm is not guaranteed to be feasible whenever $|A| \geq 4$.

\begin{table}
\centering
\caption{Infeasibility Example for DetGreedy with $|A| = 4$}
\vspace{-10pt}
\small
\begin{tabular}{|c|c|c|c|c|c|} \hline
$a_i$, $p_{a_i}$	& $a_1$, 0.4 & $a_2$, 0.4 & $a_3$, 0.1 & $a_4$, 0.1 & \multirow{3}{*}{\shortstack{\textbf{DetGreedy} \\ \textbf{chooses}}} \\ 
candidates($a_i$)		& [0.1] & [0.2] & [0.3] & [0.4] &  \\ 
\cline{2-5}
 						& \multicolumn{4}{c|}{\textbf{Min. Requirements}} & \\ \hline
Index 1	& 0 & 0 & 0 & 0 & $a_4$ \\ \hline
Index 2	& 0 & 0 & 0 & 0 & $a_3$ \\ \hline
Index 3	& 1 & 1 & 0 & 0 & \textbf{Infeasible} \\ \hline
\end{tabular}
\label{table:infeasibility_detgreedy_example}
\end{table}
\end{proof}

\subsection{Proof of Theorem \ref{th:feas2}} \label{sec:proof2}
\emph{DetConstSort algorithm is feasible per Definition~\ref{def:feasibility}}.
\begin{proof}
First, we observe that DetConstSort algorithm only includes as many candidates as needed to meet the minimum representation count requirement for each attribute value (Eq.~\ref{eq:ranking_objective_2}). Further, by position $k$ in the ranking, the algorithm would have included at most $k$ such candidates, since
\[
\sum \floor{p_{a_i} \cdot k} \leq \sum (p_{a_i} \cdot k) = (\sum p_{a_i}) \cdot k = k ~.
\]
As a result, each time a new candidate with attribute value $a_i$ has to be inserted (in order to satisfy the new minimum count requirement for $a_i$), there is sufficient empty space in the ranked list before and including $k^{th}$ position. The algorithm also never swaps a candidate to cause its attribute value $a_i$ to violate the feasibility condition (Eq.~\ref{eq:ranking_objective_2}). Since the feasibility condition is not violated at any step of the algorithm, it follows that DetConstSort is indeed feasible.
\end{proof}

\subsection{Description of Representative Ranking System Architecture}\label{sec:arch}
Figure~\ref{fig:online_architecture} details our two-tiered ranking architecture for achieving gender-representative ranking for LinkedIn Talent Search systems. There are three primary components via which we attempt to achieve representative ranking:
\vfill\eject
\begin{enumerate}
\item Computation of the gender distribution on the qualified set of candidates, along-side our first-level ranking.
\item Re-ranking of the candidate list utilizing our first-level machine-learned ranking model's scores and the desired gender distributions. The top-k$'$ candidates, ranked in a representative manner, are then sent to the second-level ranking.
\item Re-ranking of the candidate list utilizing our second-level machine learned ranking model's scores and the desired gender distributions. The top-k$''$ candidates, ranked in a representative manner, are then presented to the recruiter.
\end{enumerate}
The index corresponding to hundreds of millions of LinkedIn members is partitioned and stored in a distributed manner amongst the \emph{Searcher} nodes. Hence, we first retrieve matching candidates by issuing the search request simultaneously to multiple \emph{Searcher} nodes. We also count the number of candidates corresponding to each inferred gender amongst the matching candidates within each searcher node, and then combine these counts to get the overall gender distribution of candidates for the given search request. A top subset (k $\rightarrow$ k$'$ $\rightarrow$ k$''$) of the candidates (after ranking, and representative re-ranking) is sent to the next stage (first-level $\rightarrow$ second-level $\rightarrow$ recruiter), and finally the top k$''$ representatively ranked candidates are presented to the user (recruiter) in a page-by-page manner.

%% file: fairnessAwareRankingInSearchAndRecommendationSystems.bbl
\begin{thebibliography}{10}

\bibitem{useeoc2017}
{U.S.} equal employment opportunity commission, 2017.
\newblock \url{https://www1.eeoc.gov/eeoc/newsroom/release/1-3-17.cfm}.

\bibitem{ec2018}
European commission diversity charters, 2018.
\newblock
  \url{https://ec.europa.eu/info/strategy/justice-and-fundamental-rights/discrimination/tackling-discrimination/diversity-management/diversity-charters_en}.

\bibitem{angwin_2016}
J.~Angwin, J.~Larson, S.~Mattu, and L.~Kirchner.
\newblock Machine bias.
\newblock {\em ProPublica}, 2016.

\bibitem{arneson_2018}
R.~Arneson.
\newblock Four conceptions of equal opportunity.
\newblock {\em The Economic Journal}, 2018.

\bibitem{asudeh_2017}
A.~Asudeh, H.~V. Jagadish, J.~Stoyanovich, and G.~Das.
\newblock Designing fair ranking schemes.
\newblock In {\em SIGMOD}, 2019.

\bibitem{barocas2016big}
S.~Barocas and A.~D. Selbst.
\newblock Big data's disparate impact.
\newblock {\em California Law Review}, 104, 2016.

\bibitem{biega_2018}
A.~J. Biega, K.~P. Gummadi, and G.~Weikum.
\newblock Equity of attention: Amortizing individual fairness in rankings.
\newblock In {\em SIGIR}, 2018.

\bibitem{fairnessTutorialWWW19}
S.~Bird, B.~Hutchinson, K.~Kenthapadi, E.~Kiciman, and M.~Mitchell.
\newblock Tutorial: Fairness-aware machine learning: Practical challenges and
  lessons learned.
\newblock In {\em WWW}, 2019.
\newblock \url{https://sites.google.com/view/fairness-tutorial}.

\bibitem{bolukbasi_2016}
T.~Bolukbasi, K.-W. Chang, J.~Y. Zou, V.~Saligrama, and A.~T. Kalai.
\newblock {Man is to computer programmer as woman is to homemaker? Debiasing
  word embeddings}.
\newblock In {\em NIPS}, 2016.

\bibitem{calders_2010}
T.~Calders and S.~Verwer.
\newblock Three naive bayes approaches for discrimination-free classification.
\newblock {\em Data Mining and Knowledge Discovery}, 21(2), 2010.

\bibitem{caliskan_2017}
A.~Caliskan, J.~J. Bryson, and A.~Narayanan.
\newblock Semantics derived automatically from language corpora contain
  human-like biases.
\newblock {\em Science}, 356(6334), 2017.

\bibitem{calmon_2017}
F.~Calmon, D.~Wei, B.~Vinzamuri, K.~N. Ramamurthy, and K.~R. Varshney.
\newblock Optimized pre-processing for discrimination prevention.
\newblock In {\em NIPS}, 2017.

\bibitem{castillo2019fairness}
C.~Castillo.
\newblock Fairness and transparency in ranking.
\newblock {\em ACM SIGIR Forum}, 52(2), 2018.

\bibitem{celis_2016}
L.~E. Celis, A.~Deshpande, T.~Kathuria, and N.~K. Vishnoi.
\newblock How to be fair and diverse?
\newblock In {\em FATML}, 2016.

\bibitem{celis_2017}
L.~E. Celis, D.~Straszak, and N.~K. Vishnoi.
\newblock Ranking with fairness constraints.
\newblock In {\em ICALP}, 2018.

\bibitem{corbett_2017}
S.~Corbett-Davies, E.~Pierson, A.~Feller, S.~Goel, and A.~Huq.
\newblock Algorithmic decision making and the cost of fairness.
\newblock In {\em KDD}, 2017.

\bibitem{dwork_2012}
C.~Dwork, M.~Hardt, T.~Pitassi, and R.~Z. Omer~Reingold.
\newblock Fairness through awareness.
\newblock In {\em ITCS}, 2012.

\bibitem{friedler_2016}
S.~A. Friedler, C.~Scheidegger, and S.~Venkatasubramanian.
\newblock On the (im) possibility of fairness.
\newblock {\em arXiv:1609.07236}, 2016.

\bibitem{friedler2018comparative}
S.~A. Friedler, C.~Scheidegger, S.~Venkatasubramanian, S.~Choudhary, E.~P.
  Hamilton, and D.~Roth.
\newblock A comparative study of fairness-enhancing interventions in machine
  learning.
\newblock In {\em FAT*}, 2019.

\bibitem{friedman_1996}
B.~Friedman and H.~Nissenbaum.
\newblock Bias in computer systems.
\newblock {\em ACM Transactions on Information Systems (TOIS)}, 14(3), 1996.

\bibitem{geyik_2018}
S.~C. Geyik, Q.~Guo, B.~Hu, C.~Ozcaglar, K.~Thakkar, X.~Wu, and K.~Kenthapadi.
\newblock Talent search and recommendation systems at {LinkedIn}: Practical
  challenges and lessons learned.
\newblock In {\em SIGIR}, 2018.

\bibitem{reprank_blogpost}
S.~C. Geyik and K.~Kenthapadi.
\newblock Building representative talent search at {LinkedIn}, 2018.
\newblock LinkedIn engineering blog post,
  \url{https://engineering.linkedin.com/blog/2018/10/building-representative-talent-search-at-linkedin}.

\bibitem{hajian_2016_tutorial}
S.~Hajian, F.~Bonchi, and C.~Castillo.
\newblock Algorithmic bias: From discrimination discovery to fairness-aware
  data mining.
\newblock In {\em KDD Tutorial on Algorithmic Bias}, 2016.

\bibitem{hajian_2013}
S.~Hajian and J.~Domingo-Ferrer.
\newblock A methodology for direct and indirect discrimination prevention in
  data mining.
\newblock {\em IEEE TKDE}, 25(7), 2013.

\bibitem{hajian_2014}
S.~Hajian, J.~Domingo-Ferrer, and O.~Farr{\`a}s.
\newblock Generalization-based privacy preservation and discrimination
  prevention in data publishing and mining.
\newblock {\em Data Mining and Knowledge Discovery}, 28(5-6), 2014.

\bibitem{hardt_2016}
M.~Hardt, E.~Price, and N.~Srebro.
\newblock Equality of opportunity in supervised learning.
\newblock In {\em NIPS}, 2016.

\bibitem{jabbari_2017}
S.~Jabbari, M.~Joseph, M.~Kearns, J.~Morgenstern, and A.~Roth.
\newblock Fairness in reinforcement learning.
\newblock In {\em ICML}, 2017.

\bibitem{jarvelin_2002}
K.~Jarvelin and J.~Kekalainen.
\newblock Cumulated gain-based evaluation of {IR} techniques.
\newblock {\em ACM Trans. on Information Systems (TOIS)}, 2002.

\bibitem{joachims_2005}
T.~Joachims, L.~Granka, B.~Pan, H.~Hembrooke, and G.~Gay.
\newblock Accurately interpreting clickthrough data as implicit feedback.
\newblock In {\em SIGIR}, 2005.

\bibitem{kamiran_2010}
F.~Kamiran, T.~Calders, and M.~Pechenizkiy.
\newblock Discrimination aware decision tree learning.
\newblock In {\em ICDM}, 2010.

\bibitem{kay_2015}
M.~Kay, C.~Matuszek, and S.~A. Munson.
\newblock Unequal representation and gender stereotypes in image search results
  for occupations.
\newblock In {\em CHI}, 2015.

\bibitem{kleinberg_2017}
J.~Kleinberg, S.~Mullainathan, and M.~Raghavan.
\newblock Inherent trade-offs in the fair determination of risk scores.
\newblock In {\em ITCS}, 2017.

\bibitem{kullback_1951}
S.~Kullback and R.~A. Leibler.
\newblock On information and sufficiency.
\newblock {\em Annals of Mathematical Statistics}, 1951.

\bibitem{pedreschi_2008}
D.~Pedreschi, S.~Ruggieri, and F.~Turini.
\newblock Discrimination-aware data mining.
\newblock In {\em KDD}, 2008.

\bibitem{pedreschi_2009}
D.~Pedreschi, S.~Ruggieri, and F.~Turini.
\newblock Measuring discrimination in socially-sensitive decision records.
\newblock In {\em SDM}, 2009.

\bibitem{Sankar14}
S.~Sankar and A.~Makhani.
\newblock {Did you mean ``Galene''?}, 2014.
\newblock \url{https://engineering.linkedin.com/search/did-you-mean-galene}.

\bibitem{singh_2018}
A.~Singh and T.~Joachims.
\newblock Fairness of exposure in rankings.
\newblock In {\em KDD}, 2018.

\bibitem{verge_2010}
T.~Verge.
\newblock Gendering representation in {Spain}: Opportunities and limits of
  gender quotas.
\newblock {\em Journal of Women, Politics \& Policy}, 31(2), 2010.

\bibitem{woodworth2017learning}
B.~Woodworth, S.~Gunasekar, M.~I. Ohannessian, and N.~Srebro.
\newblock Learning non-discriminatory predictors.
\newblock In {\em COLT}, 2017.

\bibitem{yang_2017}
K.~Yang and J.~Stoyanovich.
\newblock Measuring fairness in ranked outputs.
\newblock In {\em SSDBM}, 2017.

\bibitem{zafar2017fairness}
M.~B. Zafar, I.~Valera, M.~Gomez~Rodriguez, and K.~P. Gummadi.
\newblock Fairness beyond disparate treatment \& disparate impact: Learning
  classification without disparate mistreatment.
\newblock In {\em WWW}, 2017.

\bibitem{zehlike_2017}
M.~Zehlike, F.~Bonchi, C.~Castillo, S.~Hajian, M.~Megahed, and R.~Baeza-Yates.
\newblock {FA*IR}: A fair top-k ranking algorithm.
\newblock In {\em CIKM}, 2017.

\bibitem{zemel_2013}
R.~Zemel, Y.~Wu, K.~Swersky, T.~Pitassi, and C.~Dwork.
\newblock Learning fair representations.
\newblock In {\em ICML}, 2013.

\bibitem{zliobaite2017measuring}
I.~{\v{Z}}liobait{\.e}.
\newblock Measuring discrimination in algorithmic decision making.
\newblock {\em Data Mining and Knowledge Discovery}, 31(4), 2017.

\end{thebibliography}
